Comment on
"Flow of temperature-dependent viscous fluid between parallel heated walls: Exact analytical solutions in the presence of viscous dissipation",
by K.S. Adegbie and F.I. Alao [Journal of Mathematics and Statistics, 2007, Volume 3, pp. 12-14]

Asterios Pantokratoras
Associate Professor of Fluid Mechanics
School of Engineering, Democritus University of Thrace,
67100 Xanthi – Greece
e-mail:apantokr@civil.duth.gr

In the above paper an analysis has been carried out to obtain results in the Couette flow of a Newtonian fluid with viscous dissipation and temperature dependent viscosity. The fluid viscosity is an exponential function of temperature. Exact analytical solutions are obtained for the calculation of velocity and temperature. However, there are some fundamental errors in this paper which are presented below:

1. The upper plate has temperature $T_b$ and the lower plate temperature $T_0$ ( see Figure 2.1 and equation 2.6). Taking this into account the temperature profile shown schematically in figure 2.1 is wrong.
2. The temperature boundary condition at the upper plate in equation (2.4) is wrong. The correct condition is

    $\theta(1) = 1$    (upper plate)

3. In page 13 it is mentioned that "from symmetry consideration, we need only solve the equation (3.2) with combined boundary conditions:". The above argument and the temperature boundary conditions given in equation (3.3) are all wrong. Due to unequal plate temperatures the fluid temperature profiles are non-symmetric. The correct form of the temperature profiles should be as follows ( White, 2006, page, 99)

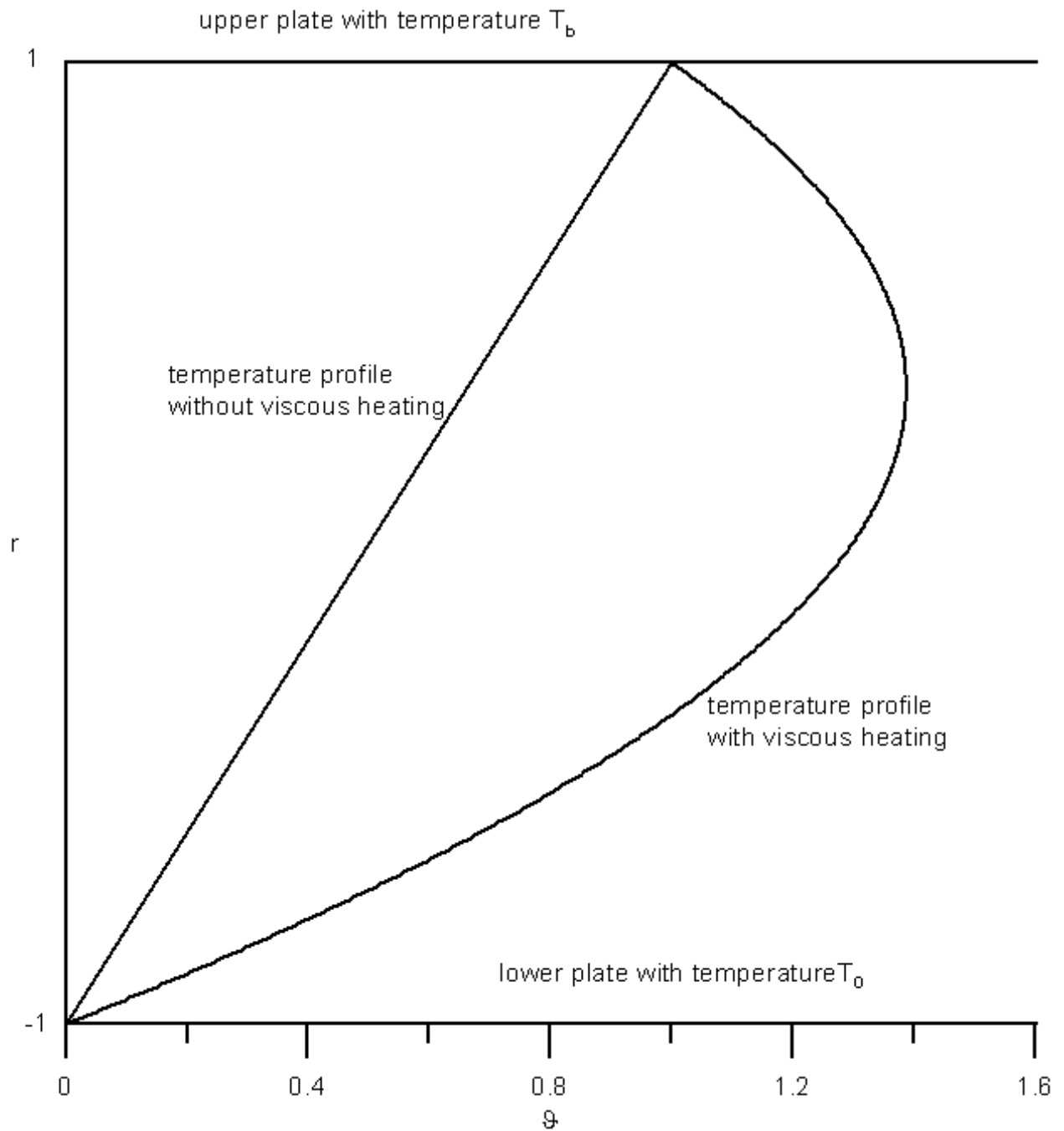

Figure 1. Temperature distribution in a Couette flow with unequal plate temperatures and viscous heating.

4. It is clear that there is no symmetry in the above temperature profile and the above paper is completely wrong.